\documentclass[prc,a4paper,twocolumn,amsmath,amssymb,showpacs,showkeys,floatfix]
{revtex4}
\usepackage{graphicx}
\usepackage{graphics}
\usepackage{latexsym}
\usepackage{booktabs}

\begin{document}

\title{Analysis of the intermediate-state contributions to neutrinoless double 
$\beta^-$ decays}
\vskip2cm
\author{Juhani Hyv\"arinen and Jouni Suhonen}
\bigskip
\affiliation{{ \small \it University of Jyvaskyla,
Department of Physics, P. O. Box 35, FI-40014, Finland.}}

\date{\today}

\begin{abstract}
A comprehensive analysis of the structure of the nuclear matrix elements (NMEs) 
of neutrinoless double beta-minus ($0\nu\beta^-\beta^-$) decays to the $0^+$ ground and
first excited states is performed in terms of the contributing multipole states in
the intermediate nuclei of $0\nu\beta^-\beta^-$ transitions. We concentrate on the transitions 
mediated by the light (l-NMEs) Majorana neutrinos. 
As nuclear model we use the proton-neutron quasiparticle random-phase 
approximation (pnQRPA) with a realistic two-nucleon interaction based on
the Bonn one-boson-exchange G matrix. In the computations we include the 
appropriate short-range correlations, nucleon form factors, higher-order nucleonic 
weak currents and restore the isospin symmetry by the isoscalar-isovector
decomposition of the particle-particle proton-neutron interaction parameter 
$g_{\rm pp}$.
\end{abstract}

\pacs{21.60.Jz, 23.40.Bw, 23.40.Hc, 27.50.+e, 27.60.+j}

\maketitle

key words: proton-neutron quasiparticle random-phase approximation,
neutrinoless double beta decays, nuclear matrix elements, light-Majorana 
exchange, isospin symmetry restoration, matrix-element
decomposition

\vskip0.5truecm

\section{Introduction}\label{intro}

Thanks to neutrino-oscillation experiments much is known about the basic 
properties of the neutrino concerning its mixing and squared mass differences.
What is not known is the absolute mass scale, the related mass hierarchy,
and the fundamental nature (Dirac or Majorana) of the neutrino. This can be
studied by analyzing the neutrinoless double beta ($0\nu2\beta$) decays of
atomic nuclei \cite{REPORT,Avignone2008,Vergados2012,Maalampi2013} through analyses
of the participating nuclear matrix elements (NMEs). The $0\nu2\beta$ decays
proceed by virtual transitions through states of all multipoles 
$J^{\pi}$ in the intermediate nucleus, $J$ being the total angular momentum 
and $\pi$ being the parity of the intermediate state. Most of the present 
interest is concentrated on the double beta-minus variant 
($0\nu\beta^-\beta^-$ decay) of the $0\nu2\beta$ decays due to their relatively 
large decay energies ($Q$ values) and natural abundancies.

In this work we concentrate on analyses of the intermediate contributions to
the $0\nu\beta^-\beta^-$ decays for the $0^+\to 0^+$ ground-state-to-ground-state and
ground-state-to-excited-state transitions in nuclear systems of experimental
interest. We focus on the light  Majorana neutrino
mediated transitions by taking into account the
appropriate short-range nucleon-nucleon correlations \cite{Kortelainen2007a},
and contributions arising from the induced currents and the finite 
nucleon size \cite{Simkovic1999}. There are several nuclear models that have recently
been used to compute the $0\nu\beta^-\beta^-$ decay NMEs (see, e.g., the extensive
discussions in
\cite{Vergados2012,Suhonen2012d,Vogel2012,Barea2013,Neacsu2014,Engel2015}).
However, the only model that avoids the closure approximation and retains the 
contributions from individual intermediate states is the proton-neutron 
quasiparticle random-phase approximation (pnQRPA) 
\cite{Suhonen2012d,Simkovic2008,Suhonen2012c,Mustonen2013,Simkovic2013}.

Some analyses of the intermediate-state contributions within the pnQRPA approach
have been performed in \cite{Simkovic2008,Mustonen2013,Rodin2006,Fang2011} 
and recently quite extensively in \cite{Hyvarinen2015,Hyvarinen2016}. 
In \cite{Hyvarinen2015} an
intermediate multipole $J^{\pi}$ decomposition was done for decays of 
$^{76}$Ge, $^{82}$Se, $^{96}$Zr, $^{100}$Mo, $^{110}$Pd,
$^{116}$Cd, $^{124}$Sn, $^{128,130}$Te, $^{136}$Xe to the ground state of the 
respective daughter nuclei, whereas in \cite{Hyvarinen2016} both an intermediate 
multipole $J^{\pi}$ decomposition and a pair-angular-momentum decomposition were done
for decays of $^{76}$Ge, $^{82}$Se, $^{96}$Zr, $^{100}$Mo, $^{110}$Pd,
$^{116}$Cd, $^{124}$Sn, $^{130}$Te, $^{136}$Xe to the lowest one or two  
excited $0^+$ states of the respective daughter nuclei. In this article we extend 
the analyses of \cite{Hyvarinen2015,Hyvarinen2016} to a more detailed
scrutiny of the intermediate contributions to the $0\nu\beta^-\beta^-$ decay NMEs
of the above-mentioned nuclei.

\section{Theory background}\label{sec:formalism}

In this section a very brief introduction to the computational framework of the
present calculations is given. The present analyses are based on the calculations
done in Refs.~\cite{Hyvarinen2015,Hyvarinen2016} and more details on the used
theory tools can be checked there and in \cite{Hyvarinen2015b}. We assume here
that the $0\nu\beta^-\beta^-$ decay proceeds via the light Majorana neutrino
so that the inverse half-life can be written as
\begin{equation} \label{eq:half-life}
\left\lbrack t_{1/2}^{(0 \nu)}(0_i^+ \rightarrow 0_f^+)\right\rbrack^{-1}
 = g_{\rm A}^4G_{0\nu}\left\vert M^{(0\nu)}\right\vert^2  |\left\langle m_{\nu} \right\rangle|^2
 \, ,
\end{equation}
where $G_{0\nu}$ is a phase-space factor for the final-state leptons defined
here without the axial-vector coupling constant $g_{\rm A}$. The 
quantity \( \left\langle m_{\nu} \right\rangle \)
denotes the neutrino effective mass and describes the physics beyond the standard model
\cite{Hyvarinen2015}. The
quantity $M^{(0\nu)}$ is the light-neutrino nuclear matrix element (l-NME). The nuclear
matrix element can be
decomposed into Gamow-Teller (GT), Fermi (F) and tensor (T) contributions as
\begin{equation} \label{eq:0nume}
M^{(0\nu)} = M_{\rm GT}^{(0\nu)} - \left( \frac{g_{\rm V}}{g_{\rm A}}
\right)^{2} M_{\rm F}^{(0\nu)}+ M_{\rm T}^{(0\nu)} \, ,
\end{equation}
where $g_{\rm V}$ is the vector coupling constant.
 
Each of the NMEs $K=\textrm{GT,F,T}$ in (\ref{eq:0nume}) can be decomposed in terms
of the intermediate multipole contributions $J^{\pi}$ as
\begin{equation} \label{eq:decomp}
M_{K}^{(0\nu)} = \sum_{J^{\pi}}M_{K}^{(0\nu)}(J^{\pi}) \,,
\end{equation}
where each multipole contribution is, in turn, decomposed in terms of the
two-particle transition matrix elements and one-body transition densities as
\begin{align} \label{eq:NMEs}
M_{K}^{(0\nu)}&(J^{\pi})  =  \sum_{k_{1},k_{2},J'} \sum_{pp'nn'}
(-1)^{j_{n}+j_{p'}+J+J'} \sqrt{2J'+1} \nonumber \\
&  \times \left\{ \begin{array}{ccc} j_{p} & j_{n} & J \\
j_{n'} & j_{p'} & J'\end{array} \right\} 
( pp':J' \vert\vert  {\mathcal O}_K \vert\vert nn':J' )
\\ &  \times
( 0^{+}_{f} \vert \vert  \left[ c^{\dag}_{p'}
\tilde{c}_{n'}\right]_J \vert \vert J^{\pi}_{k_{1}} )
\langle J^{\pi}_{k_{1}} \vert J^{\pi}_{k_{2}} \rangle
( J^{\pi}_{k_{2}} \vert \vert  \left[ c^{\dag}_{p}
\tilde{c}_{n}\right]_J \vert \vert 0^{+}_{i}) \, , \nonumber
\end{align}
where $k_1$ and $k_2$ label the different pnQRPA solutions for a given
multipole $J^{\pi}$ and the indices $p,p',n,n'$ denote the proton and neutron 
single-particle quantum numbers. The operators ${\mathcal O}_K$ inside the
two-particle matrix element contain the neutrino potentials for the light
Majorana neutrinos, the characteristic two-particle operators for the 
different $K=\textrm{GT,F,T}$ and a function taking into account the
short-range correlations (SRC) between the two decaying neutrons in the mother 
nucleus of $0\nu\beta^-\beta^-$ decay \cite{Hyvarinen2015}. The final $0^+$ state,
$0^{+}_{f}$, can be either the ground state or an excited state of the 
$0\nu\beta^-\beta^-$ daugter nucleus, and the overlap factor between the two
one-body transition densities helps connect the corresponding intermediate $J^{\pi}$
states emerging from the pnQRPA calculations in the mother and daughter nuclei.
The one-body transition densities are exposed in detail in the articles 
\cite{Hyvarinen2015,Hyvarinen2016}.

As mentioned before, our calculations contain the appropriate short-range
correlators, nucleon form factors and higher-order nucleonic weak currents.
In addition, we decompose the particle-particle proton-neutron interaction strength
parameter $g_{\rm pp}$ of the pnQRPA into its isoscalar ($T=0$) and isovector 
($T=1$) components and adjust these components independently as described in
\cite{Hyvarinen2015}: The isovector component is fixed such that the NME of the 
two-neutrino double beta decay ($2\nu\beta^-\beta^-$) vanishes and the isospin 
symmetry is thus restored for both the $2\nu\beta^-\beta^-$ and 
$0\nu\beta^-\beta^-$ decays. The isoscalar component, in turn, is fixed such that
the measured half-life of the $2\nu\beta^-\beta^-$ decay is reproduced. The
resulting values of both components of $g_{\rm pp}$ are shown in Table~I of 
Ref.~\cite{Hyvarinen2015}. The details of the chosen valence spaces and the determination
of the other hamiltonian parameters are presented in 
\cite{Hyvarinen2015,Hyvarinen2016}.

\section{Results and Discussion}\label{sec:results}

In this section we discuss and present the results of our calculations. Presentation
of the results follows top to bottom approach. First we analyze the multipole decompositions
and total cumulative sums of the matrix elements. From these we can extract the most important
multipole components and energy regions contibuting to the NMEs. After this we continue and
dissect the most important multipole components into contributions coming 
from different individual
states of the \(0\nu\beta\beta\) intermediate nucleus. Throughout these computations we have 
used a conservatively quenched value of the axial vector coupling \(g_\mathrm{A} = 1.00\).

\subsection{Ground-state-to-ground-state transitions}

\begin{figure*}[htbp]
\includegraphics*[scale=0.6]{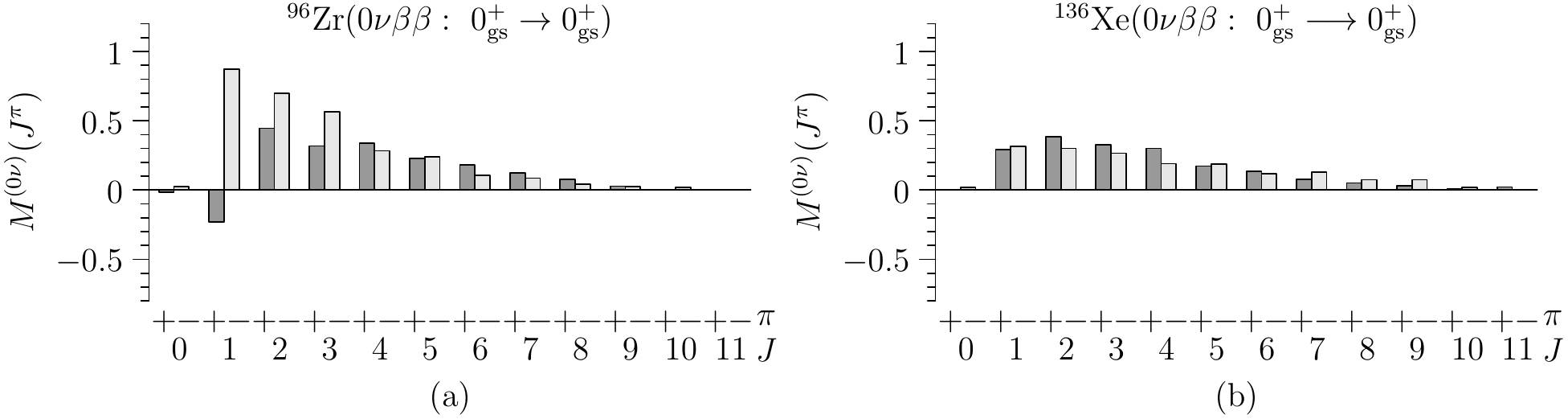}
\caption{Multipole decomposition of the l-NME for the nuclei 
\(^{96}\)Zr and \(^{136}\)Xe corresponding
to the \( 0^+_\mathrm{gs} \rightarrow 0^+_\mathrm{gs}\) decay transitions.}  
\label{fig:rpa-decomp-96-136-gs}
\end{figure*}

Let us begin by considering the ground-state-to-ground-state decays mediated by
light neutrino exchange. In Fig. \ref{fig:rpa-decomp-96-136-gs} (a)-(b) we have 
plotted the multipole decomposition (\ref{eq:decomp})
of the l-NMEs corresponding to the \(A = 96\) and 136 nuclear systems.
For most nuclei considered in this work, the leading multipole component is \(1^-\) This is the
case also for the nucleus \(^{96}\)Zr shown in Fig. \ref{fig:rpa-decomp-96-136-gs} (a). 
Most important contribution
to the NMEs comes from the lowest multipole components \(1^{\pm}-4^{\pm}\).
It can also be observed that the shape of the overall multipole distribution is
leveled when going towards heavier nuclei. This can be seen by comparing the
distribution of \(^{96}\)Zr with the distribution of \(^{136}\)Xe
displayed in Fig. \ref{fig:rpa-decomp-96-136-gs} (b).

\begin{figure*}[htbp]
\includegraphics*[scale=0.6]{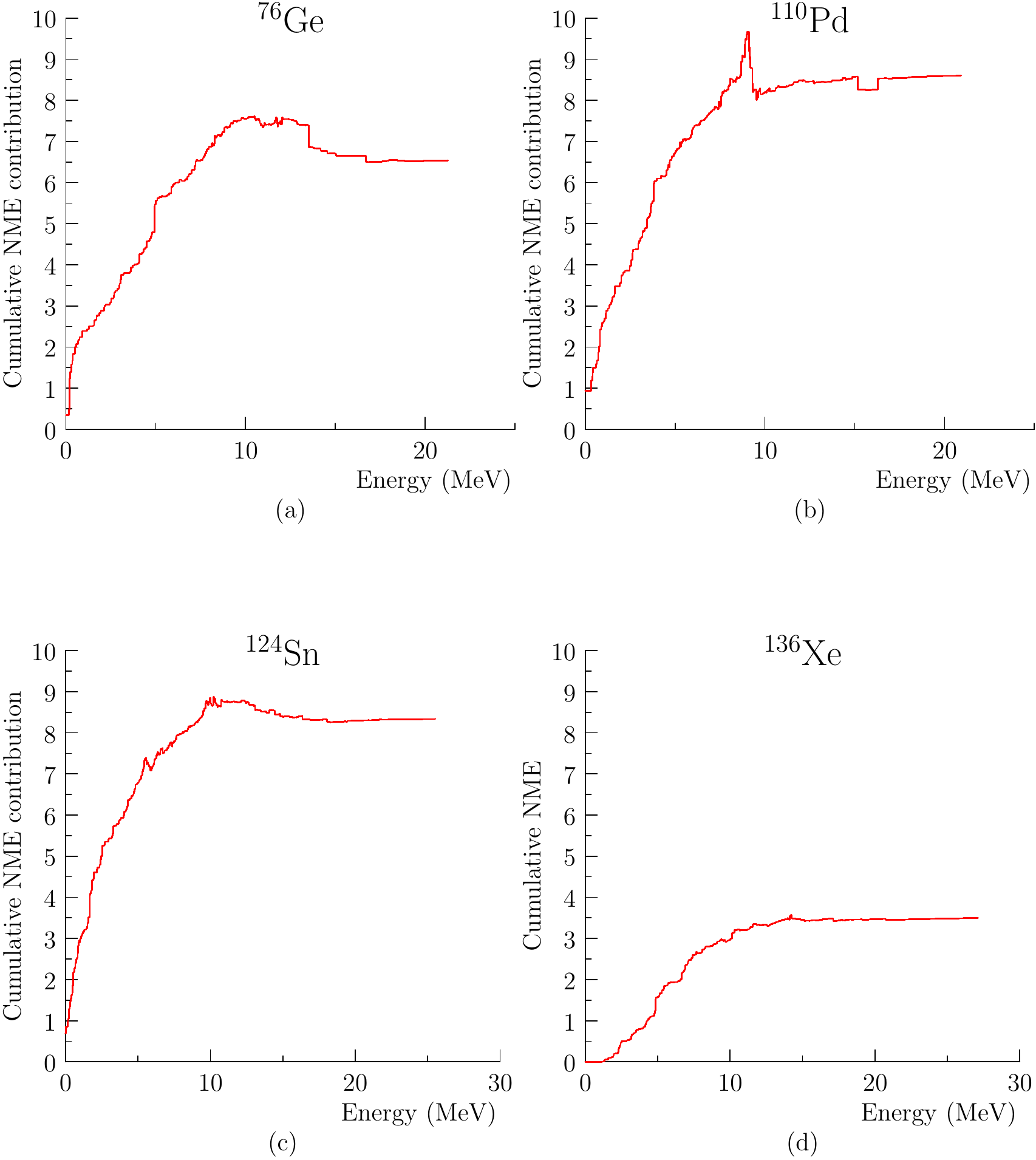}
\caption{Cumulative values of the computed l-NMEs corresponding to the \(0^+_\mathrm{gs} 
\rightarrow 0^+_\mathrm{gs} \) decay transitions for the nuclear systems
\(A = 76, 110, 124\) and 136. The horizontal axis gives the excitation
energies of the intermediate states contributing to the \(0\nu\beta\beta\)
transition.}  \label{fig:cumsumL-76-110-124-136}
\end{figure*}

Nuclei can be grouped into different types according to the shapes
of their cumulative NME distributions. For \(0^+_\mathrm{gs} \longrightarrow 0^+_\mathrm{gs} \)
transitions via light neutrino exchange, we can differentiate four types of nuclei. 
\textbf{Type 1}: 
Nuclei belonging to this type are \(^{76}\)Ge, \(^{82}\)Se, \(^{96}\)Zr and \(^{128}\)Te.
Representative of this type, \(^{76}\)Ge, is presented in Fig. 
\ref{fig:cumsumL-76-110-124-136}, panel (a). Characteristic feature of 
the cumulative sum distribution belonging to type 1 is the strong drop in the value of the
NME occuring between 12-17 MeV. Soon after this drop the NME saturates as can be seen from
panel (a).
\textbf{Type 2}: Nuclei belonging to this type are \(^{100}\)Mo and \(^{110}\)Pd.
Representative of this type, \(^{110}\)Pd, is presented in Fig. 
\ref{fig:cumsumL-76-110-124-136}, panel (b). Characteristic
feature of this type is the large enhancement and almost immediate cancellation of this
enhancement around 10 MeV. This produces a spike like structure into the cumulative sum
distribution as can be seen from panel (b).  
\textbf{Type 3}: Nuclei belonging to type 3 are \(^{116}\)Cd, \(^{124}\)Sn and \(^{130}\)Te.
Type 3 is represented by \(^{124}\)Sn, shown in Fig. 
\ref{fig:cumsumL-76-110-124-136}, panel (c). Characteristic features of this type
are that there occurs neither sharp cancellation of the NME around 12-17 MeV, as in type 1,
nor a spike like structure around 10 MeV, as in type 2. Value of the NME rather increases
more or less smoothly to its highest value and then smoothly saturates to its final value 
around 20 MeV.
\textbf{Type 4}: Type 4 is special in a sence that it icludes only one nucleus, \(^{136}\)Xe.
Cumulative sum of the NME for \(^{136}\)Xe is shown in Fig. 
\ref{fig:cumsumL-76-110-124-136}, panel (d). Characteristic
feature of type 4 is that the lowest energy region, roughly between 0-1.5 MeV, 
contributes practically nothing
to the value of the NME as can be noticed from panel (d).  

\begin{table*}[]
\caption{Most important multipoles and intermediate states contributing
to the ground-state-to-ground-state \(0\nu\beta\beta\) decays mediated by
the light neutrino exchange. Columns \(E\) give the energies (in MeVs) and multipoles of the
intermediate states. Multipoles are organized from left to right in terms of their
importance, the most important being on the left. Columns labeled C give the corresponding NME contributions.
Last two numbers in each C column give the summed contribution and the percentual
part which the displayed states give to the total multipole strength.} 
\label{tab:gs1}
\begin{ruledtabular}       
\begin{tabular}{ l c c c c c c c c c c}
Nucleus & \(E(2^-)\) &  C  & \(E(1^-)\) &  C  & \(E(2^+)\) &  C  & \(E(1^+)\) &  C  &  
\(E(3^-)\) & C   \\
\(^{76}\)Ge&	0.22	&	0.748	&	5.87	&	0.155	&	0.51	
&	0.166	&	0.00	&	0.344	&	0.30	&	0.189	\\
&		&		&	6.32	&	0.058	&	1.87	&	
0.056	&	4.09	&	0.199	&	0.92	&	0.148	\\
&		&		&	7.00	&	0.077	&	3.04	&	
0.094	&	4.48	&	0.148	&	6.77	&	0.056	\\
&		&		&	7.16	&	0.064	&	3.62	&	
0.108	&	4.94	&	0.578	&	6.85	&	0.053	\\
&		&		&	8.27	&	0.190	&	4.81	&	
0.058	&	10.80	&	-0.053	&	11.63	&	0.062	\\
&		&		&	11.04	&	0.084	&	7.73	&	
0.054	&	11.75	&	-0.109	&	12.07	&	-0.053	\\
&		&		&	12.03	&	0.165	&		&		
&	13.52	&	-0.522	&		&		\\
&		&		&	16.70	&	0.158	&		&		
&		&		&		&		\\
&		&	0.748	&		&	0.635	&		&	
0.556	&		&	0.668	&		&	0.454	\\
&		&	79\%	&		&	83\%	&		&	
77\%	&		&	88\%	&		&	70\%	\\
\hline
        & \(E(2^-)\) &  C  & \(E(2^+)\) &  C  & \(E(1^-)\) &  C  & \(E(3^-)\) & C   &  
\(E(1^+)\) &  C  \\
\(^{82}\)Se&	0.00	&	0.510	&	0.65	&	0.137	&	5.27	
&	0.116	&	0.07	&	0.140	&	0.19	&	0.264	\\
&		&		&	1.73	&	0.065	&	6.85	&	
0.065	&	0.82	&	0.138	&	3.16	&	0.095	\\
&		&		&	2.22	&	0.051	&	7.98	&	
0.134	&		&		&	4.07	&	0.065	\\
&		&		&	3.56	&	0.084	&	9.79	&	
0.083	&		&		&	4.55	&	0.105	\\
&		&		&	4.05	&	0.052	&	12.25	&	
0.065	&		&		&	5.32	&	0.559	\\
&		&		&	4.93	&	0.054	&	17.41	&	
0.070	&		&		&	7.01	&	-0.253	\\
&		&		&		&		&		&		
&		&		&	14.53	&	-0.396	\\
&		&	0.510	&		&	0.442	&		&	
0.393	&		&	0.278	&		&	0.439	\\
&		&	81\%	&		&	81\%	&		&	76\%	
&		&	57\%	&		&	92\%	\\
\hline
        & \(E(1^-)\) & C   & \(E(2^-)\) &  C  & \(E(3^-)\) & C   & \(E(2^+)\) &  C  
&  \(E(4^+)\) & C   \\
\(^{96}\)Zr&	1.75	&	0.056	&	0.92	&	0.498	&	1.35	
&	0.151	&	0.64	&	0.150	&	1.05	&	0.099	\\
&	2.28	&	0.063	&	2.21	&	0.065	&	2.35	&	
0.051	&	1.63	&	0.114	&	1.68	&	0.071	\\
&	2.52	&	0.150	&	3.75	&	0.050	&	7.77	&	
0.064	&		&		&	5.69	&	0.064	\\
&	4.46	&	0.050	&	4.43	&	0.052	&	11.36	&	
-0.091	&		&		&		&		\\
&	5.04	&	0.077	&	8.53	&	0.057	&		&		
&		&		&		&		\\
&	5.27	&	0.209	&	8.77	&	-0.056	&		&		
&		&		&		&		\\
&	8.65	&	0.060	&		&		&		&		
&		&		&		&		\\
&	11.33	&	0.061	&		&		&		&		
&		&		&		&		\\
&		&	0.728	&		&	0.666	&		&	0.175	
&		&	0.265	&		&	0.234	\\
&		&	83\%	&		&	96\%	&		&	31\%	
&		&	59\%	&		&	69\%	\\
\hline
        & \(E(1^-)\) &  C  & \(E(2^+)\) & C   & \(E(4^+)\) & C   & \(E(3^+)\) & C   &  \(E(2^-)\) 
&  C  \\
\(^{100}\)Mo&	3.12	&	0.307	&	0.90	&	0.192	&	1.32	
&	0.158	&	1.33	&	0.202	&	1.76	&	0.205	\\
&	4.41	&	0.109	&	2.10	&	0.154	&	2.26	&	
0.097	&	1.68	&	0.076	&	2.85	&	0.147	\\
&	6.68	&	0.091	&	11.68	&	0.082	&		&		
&	7.43	&	0.152	&	5.52	&	-0.061	\\
&	11.15	&	0.117	&	11.97	&	-0.070	&		&		
&	7.76	&	-0.173	&	10.17	&	0.219	\\
&	16.72	&	0.058	&		&		&		&		
&		&		&	10.93	&	-0.194	\\
&	20.26	&	-0.060	&		&		&		&		
&		&		&		&		\\
&	23.90	&	-0.060	&		&		&		&		
&		&		&		&		\\
&		&	0.562	&		&	0.359	&		&	0.255	
&		&	0.257	&		&	0.316	\\
&		&	70\%	&		&	62\%	&		&	53\%	
&		&	54\%	&		&	72\%	\\        
\end{tabular}
\end{ruledtabular}
\end{table*}
\begin{table*}
\caption{Most important multipoles and intermediate states contributing
to the ground-state-to-ground-state \(0\nu\beta\beta\) decays mediated by
the light neutrino exchange. Columns \(E\) give the energies (in MeVs) and multipoles of the
intermediate states. Multipoles are organized from left to right in terms of their
importance, the most important being on the left. Columns labeled C give the corresponding NME contributions.
Last two numbers in each C column give the summed contribution and the percentual
part which the displayed states give to the total multipole strength.}
\label{tab:gs2}
\begin{ruledtabular}         
\begin{tabular}{lcccccccccc}
Nucleus & \(E(1^-)\) &  C  & \(E(2^-)\) &  C  & \(E(1^+)\) &  C  & \(E(3^-)\) &  C  &  \(E(2^+)\) & C   \\
\(^{110}\)Pd&	2.95	&	0.130	&	0.82	&	0.387	&	0.00	&	0.938	&	1.14	&	0.118	&	0.33	&	0.244	\\
&	3.19	&	0.106	&	2.47	&	0.091	&	4.70	&	-0.062	&	1.64	&	0.128	&	0.95	&	0.087	\\
&	3.44	&	0.221	&	2.63	&	0.163	&	9.61	&	0.153	&	1.91	&	0.080	&	8.68	&	0.061	\\
&	3.81	&	0.426	&	5.94	&	0.107	&	9.75	&	-0.146	&	3.07	&	0.053	&	8.73	&	0.075	\\
&	4.52	&	0.126	&	8.88	&	0.161	&	10.22	&	0.079	&	3.61	&	0.075	&	9.00	&	0.061	\\
&	9.11	&	0.109	&	9.54	&	-0.256	&	15.16	&	-0.316	&	5.33	&	0.066	&	9.07	&	0.053	\\
&		&		&		&		&	16.30	&	0.256	&	8.08	&	0.167	&	9.16	&	0.199	\\
&		&		&		&		&		&		&	8.37	&	-0.051	&	9.30	&	-0.366	\\
&		&		&		&		&		&		&	8.45	&	0.052	&		&		\\
&		&		&		&		&		&		&	8.68	&	0.275	&		&		\\
&		&		&		&		&		&		&	9.11	&	-0.468	&		&		\\
&		&	1.118	&		&	0.653	&		&	0.903	&		&	0.496	&		&	0.414	\\
&		&	77\%	&		&	68\%	&		&	96\%	&		&	60\%	&		&	55\%	\\
\hline
        & \(E(1^-)\) &  C  & \(E(3^-)\) &  C  & \(E(2^-)\) &  C  & \(E(1^+)\) &  C  &  \(E(3^+)\) & C   \\
\(^{116}\)Cd&	3.61	&	0.223	&	1.72	&	0.081	&	1.84	&	0.065	&	0.00	&	0.378	&	0.90	&	0.158	\\
&	4.47	&	0.099	&	2.30	&	0.053	&	2.93	&	0.287	&	8.37	&	-0.102	&	1.40	&	0.068	\\
&	5.37	&	0.118	&	2.79	&	0.051	&	7.56	&	0.075	&	9.51	&	0.087	&	3.89	&	0.057	\\
&	5.87	&	0.140	&	7.24	&	0.104	&	8.31	&	0.246	&	10.74	&	-0.083	&		&		\\
&	8.61	&	0.102	&		&		&	8.34	&	0.103	&	11.20	&	0.081	&		&		\\
&	23.07	&	0.071	&		&		&	8.45	&	-0.262	&	13.75	&	0.080	&		&		\\
&		&		&		&		&	9.69	&	-0.051	&	13.79	&	-0.087	&		&		\\
&		&		&		&		&		&		&	15.73	&	-0.363	&		&		\\
&		&		&		&		&		&		&	15.83	&	0.440	&		&		\\
&		&		&		&		&		&		&	16.51	&	-0.082	&		&		\\
&		&		&		&		&		&		&	16.83	&	0.092	&		&		\\
&		&	0.753	&		&	0.289	&		&	0.463	&		&	0.439	&		&	0.284	\\
&		&	72\%	&		&	65\%	&		&	106\%	&		&	102\%	&		&	76\%	\\          
\hline
        & \(E(1^-)\) &  C  & \(E(1^+)\) &  C  & \(E(2^+)\) &  C  & \(E(2^-)\) &  C  &  \(E(3^-)\) & C   \\
\(^{124}\)Sn&	1.68	&	0.522	&	0.00	&	0.690	&	0.23	&	0.157	&	0.52	&	0.271	&	0.36	&	0.050	\\
&	4.82	&	0.089	&	1.00	&	-0.067	&	0.60	&	0.083	&	1.82	&	0.225	&	0.49	&	0.110	\\
&	6.54	&	0.082	&	2.56	&	0.252	&	1.06	&	0.056	&	4.55	&	0.051	&	1.95	&	0.150	\\
&	9.57	&	0.059	&	3.31	&	0.153	&	2.15	&	0.066	&	7.60	&	0.057	&	9.67	&	0.086	\\
&	10.74	&	0.159	&	6.72	&	-0.130	&	7.11	&	0.055	&	7.66	&	-0.064	&	9.70	&	0.062	\\
&	14.07	&	0.065	&	9.51	&	0.098	&		&		&	10.20	&	0.183	&	9.82	&	-0.087	\\
&	14.46	&	-0.099	&	13.09	&	-0.112	&		&		&	10.35	&	-0.094	&	12.64	&	-0.053	\\
&	14.83	&	-0.058	&		&		&		&		&		&		&		&		\\
&	16.36	&	-0.087	&		&		&		&		&		&		&		&		\\
&	18.05	&	-0.067	&		&		&		&		&		&		&		&		\\
&		&	0.664	&		&	0.884	&		&	0.417	&		&	0.629	&		&	0.318	\\
&		&	55\%	&		&	95\%	&		&	53\%	&		&	89\%	&		&	49\%	\\
\end{tabular}
\end{ruledtabular}
\end{table*}
\begin{table*}[]
\caption{Most important multipoles and intermediate states contributing
to the ground-state-to-ground-state \(0\nu\beta\beta\) decays mediated by
the light neutrino exchange. Columns \(E\) give the energies (in MeVs) and multipoles of the
intermediate states. Multipoles are organized from left to right in terms of their
importance, the most important being on the left. Columns labeled C give the corresponding NME contributions.
Last two numbers in each C column give the summed contribution and the percentual
part which the displayed states give to the total multipole strength.}
\label{tab:gs3}
\begin{ruledtabular}         
\begin{tabular}{lcccccccccc}
Nucleus & \(E(1^-)\) &  C  & \(E(2^+)\) &  C  & \(E(3^-)\) &  C  & \(E(2^-)\) &  C  &  \(E(3^+)\) & C   \\
\(^{128}\)Te&	4.22	&	0.200	&	0.04	&	0.066	&	0.16	&	0.055	&	0.61	&	0.335	&	0.02	&	0.074	\\
&	4.72	&	0.060	&	0.51	&	0.052	&	0.58	&	0.140	&	4.02	&	0.060	&	2.37	&	0.063	\\
&	6.21	&	0.078	&	2.93	&	0.084	&	3.97	&	0.052	&	4.55	&	0.101	&	6.22	&	0.078	\\
&	6.44	&	0.059	&	3.97	&	0.053	&	10.04	&	0.061	&	4.89	&	-0.070	&	6.77	&	-0.065	\\
&	8.07	&	-0.084	&	6.77	&	0.050	&		&		&	10.14	&	0.136	&	9.82	&	-0.062	\\
&	8.30	&	0.151	&		&		&		&		&	10.57	&	-0.056	&	10.27	&	-0.057	\\
&	8.98	&	0.068	&		&		&		&		&	11.55	&	0.058	&		&		\\
&	10.69	&	-0.052	&		&		&		&		&		&		&		&		\\
&	11.12	&	0.182	&		&		&		&		&		&		&		&		\\
&	17.48	&	-0.079	&		&		&		&		&		&		&		&		\\
&	19.19	&	-0.100	&		&		&		&		&		&		&		&		\\
&		&	0.484	&		&	0.305	&		&	0.308	&		&	0.564	&		&	0.154	\\
&		&	69\%	&		&	52\%	&		&	58\%	&		&	120\%	&		&	33\%	\\
\hline
Nucleus & \(E(1^-)\) &  C  & \(E(2^+)\) &  C  & \(E(3^-)\) &  C  & \(E(3^+)\) &  C  &  \(E(2^-)\) & C   \\
\(^{130}\)Te&	4.18	&	0.184	&	0.13	&	0.054	&	0.84	&	0.113	&	0.10	&	0.056	&	0.97	&	0.277	\\
&	5.72	&	0.059	&	3.16	&	0.089	&		&		&	0.36	&	0.052	&	10.25	&	0.061	\\
&	6.27	&	0.059	&	4.70	&	0.064	&		&		&	2.60	&	0.056	&	11.33	&	-0.100	\\
&	8.56	&	0.096	&	10.49	&	-0.065	&		&		&	6.35	&	0.063	&		&		\\
&	11.26	&	0.115	&	10.57	&	0.145	&		&		&	6.83	&	0.054	&		&		\\
&	17.77	&	-0.062	&	10.97	&	-0.088	&		&		&	12.31	&	0.059	&		&		\\
&	19.51	&	-0.079	&	16.44	&	0.150	&		&		&	12.41	&	-0.065	&		&		\\
&		&		&	16.54	&	-0.132	&		&		&		&		&		&		\\
&		&	0.373	&		&	0.216	&		&	0.113	&		&	0.166	&		&	0.237	\\
&		&	61\%	&		&	40\%	&		&	25\%	&		&	39\%	&		&	56\%	\\
\hline
Nucleus & \(E(2^+)\) &  C  & \(E(3^+)\) &  C  & \(E(1^-)\) &  C  & \(E(2^-)\) &  C  &  \(E(4^+)\) & C   \\
\(^{136}\)Xe&	1.59	&	0.033	&	1.35	&	0.033	&	6.65	&	0.141	&	4.86	&	0.188	&	2.34	&	0.032	\\
&	2.26	&	0.038	&	2.39	&	0.054	&	7.32	&	0.044	&	7.42	&	0.068	&	2.44	&	0.061	\\
&	2.50	&	0.037	&	4.81	&	0.050	&	10.31	&	0.053	&		&		&	4.43	&	0.031	\\
&	5.29	&	0.067	&	8.27	&	0.035	&		&		&		&		&	4.76	&	0.049	\\
&	6.57	&	0.033	&	9.99	&	0.027	&		&		&		&		&	7.67	&	0.034	\\
&	7.05	&	0.027	&	10.65	&	-0.027	&		&		&		&		&	9.10	&	0.045	\\
&	14.11	&	0.087	&		&		&		&		&		&		&	9.65	&	-0.038	\\
&	14.25	&	-0.085	&		&		&		&		&		&		&		&		\\
&		&	0.238	&		&	0.172	&		&	0.238	&		&	0.256	&		&	0.214	\\
&		&	62\%	&		&	53\%	&		&	76\%	&		&	85\%	&		&	71\%	\\  
\end{tabular}
\end{ruledtabular}
\end{table*}
Using the multipole decompositions, we have extracted the most important
multipole components contibuting to the light neutrino mediated 
ground-state-to-ground-state decays. These most important components
can be divided into contributions coming from different energy levels
of the \(0\nu\beta\beta\) intermediate nucleus. These contributions
are collected into Table \ref{tab:gs1} for \(A = 76-100 \) systems, into Table
\ref{tab:gs2} for \(A = 110-124 \) systems and into Table \ref{tab:gs3} for \(A = 128-136 \)
systems.  We see from the tables that often a very small set of states
collects the largest part of a given multipole contribution to the
NMEs. Also in some cases notable contributions are coming from high excitation energies, well
above 10 MeV, like in the case of $1^-$ contributions for almost all nuclei,
$1^+$ contributions for \(^{76}\)Ge, \(^{82}\)Se, \(^{110}\)Pd, \(^{116}\)Cd and \(^{124}\)Sn, 
$2^+$ contributions for \(^{130}\)Te and \(^{136}\)Xe and a $3^-$ contribution for \(^{124}\)Sn.

We notice a single-state dominance for the \(2^-\) mode in nuclei
\(^{76}\)Ge, \(^{82}\)Se and \(^{96}\)Zr. In Ref. \cite{Ejiri2014} an analysis
of the unique first forbidden single \(\beta^{\pm} \) \( 2^- \rightarrow 0^+ \)
ground-state-to-ground-state transitions in the mass region
\(A = 72-132\) was performed. It was
found that a strong renormalization of the axial vector \(2^-\)
single \(\beta\) matrix elements is needed to be able to explain 
the experimental transition rates. It was then speculated that 
a same kind of an effect may also appear in the \(0\nu\beta\beta\)
NMEs. This may have a large effect on the \(0\nu\beta\beta\) transition rates 
due to the important contribution of the \(2^-\) multipole to the \(0\nu\beta\beta\) NMEs.

\subsection{Ground-state-to-excited-state decays}

\begin{figure*}[htbp]
\includegraphics*[scale=0.6]{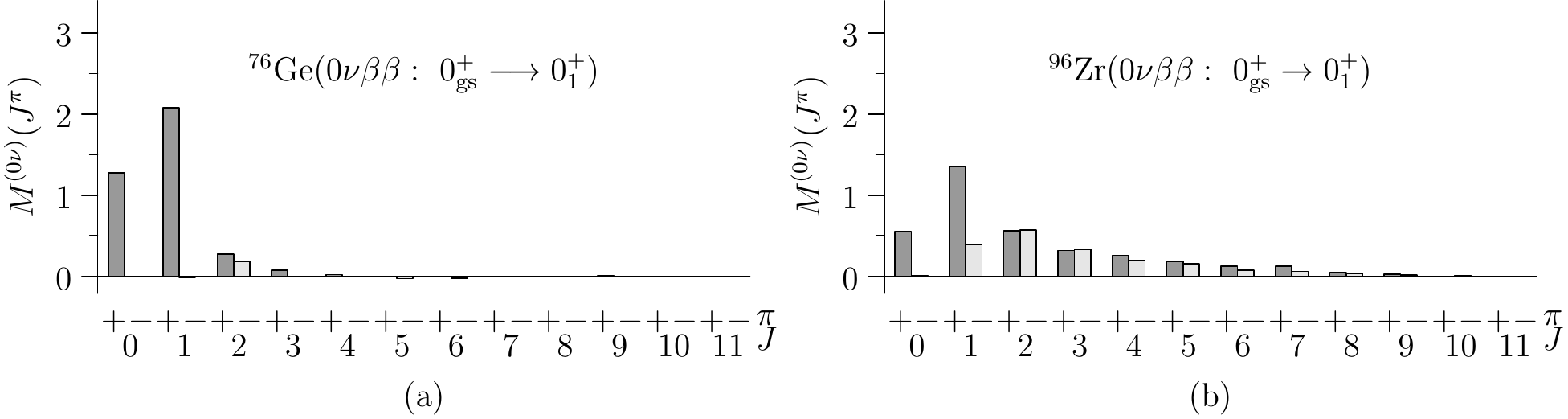}
\caption{Multipole decomposition of the l-NME for the nuclei \(^{76}\)Ge and \(^{96}\)Zr corresponding
to the \( 0^+_\mathrm{gs} \rightarrow 0^+_1\) decay transitions.}  \label{fig:rpa-exc1-decomp-76-96}
\end{figure*}

Let us then consider \(0^+_\mathrm{gs} \rightarrow 0^+_1 \) transitions mediated by the
light neutrino exchange. In Fig. \ref{fig:rpa-exc1-decomp-76-96} (a)-(b) we have 
plotted the multipole decomposition of the l-NMEs corresponding to the \(A = 76\) 
and 96 nuclear systems. 
The multipole distributions for the excited-state transitions are greatly different from those
corresponding to the ground-state transitions. Usually there is only a couple of 
multipoles, \(0^+\) and \(1^+\),
which give by far the largest contribution to the NMEs. In this sense the excited-state transitions
are more simple than the ground-state transitions. Typical example is the nucleus \(^{76}\)Ge, 
displayed in Fig. \ref{fig:rpa-exc1-decomp-76-96} (a). One nucleus deviating from this 
trend is \(^{96}\)Zr which is presented in
Fig. \ref{fig:rpa-exc1-decomp-76-96} (b). Its multipole distribution resembles somewhat more 
those shown for the ground-state decays
in Fig. \ref{fig:rpa-decomp-96-136-gs} (a)-(b). 

\begin{figure*}[htbp]
\includegraphics*[scale=0.6]{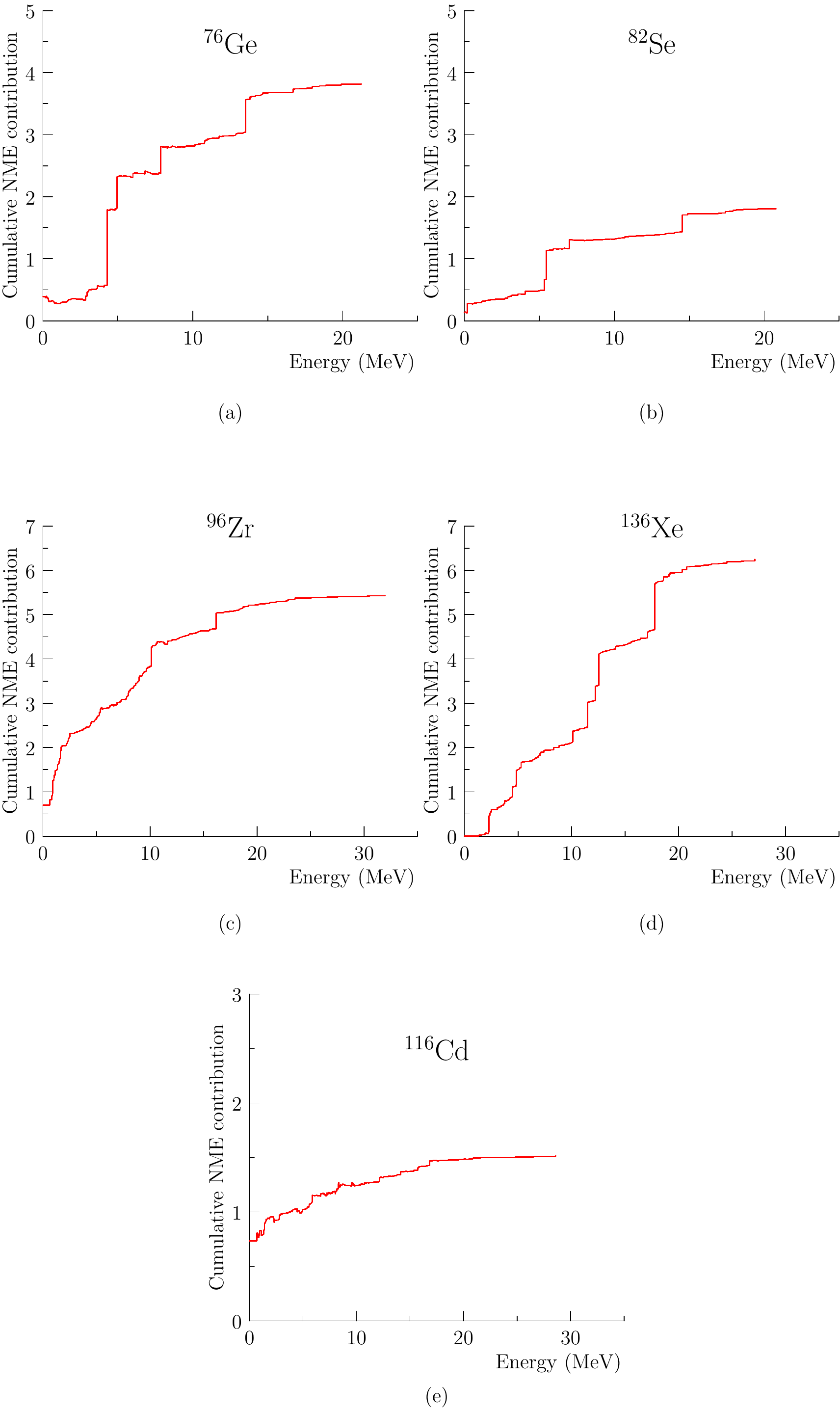}
\caption{Cumulative values of the computed l-NMEs corresponding to the \(0^+_\mathrm{gs} 
\rightarrow 0^+_1 \) decay transitions for the nuclear systems
\(A = 76, 82, 96, 116\) and 136. The horizontal axis gives the excitation
energies of the intermediate states contributing to the \(0\nu\beta\beta\)
transition.}  \label{fig:cumsumL-exc1}
\end{figure*}

Again we can divide nuclei into different groups by considering the shapes of their
total cumulative sum distributions. For \(0^+_\mathrm{gs} \longrightarrow 0^+_1 \)
transitions via light neutrino exchange, we can differentiate two types of nuclei.
\textbf{Type 1}: Nuclei belonging to type 1 are \(^{76}\)Ge, \(^{82}\)Se, \(^{124}\)Sn,
\(^{130}\)Te and \(^{136}\)Xe. Typical examples of this type, \(^{76}\)Ge, \(^{82}\)Se,
\(^{136}\)Xe, are shown in Fig. \ref{fig:cumsumL-exc1}, panels (a), (b) and (d). 
Characteristic feature of this type is that there exist only few energy states 
which give most of the total matrix element producing a staircase like structure as seen in
the panels. For example, for \(^{76}\)Ge there seems to be only five such energy states.
\textbf{Type 2}: Nuclei belonging to this type are \(^{96}\)Zr, \(^{100}\)Mo, \(^{110}\)Pd 
and \(^{116}\)Cd. Typical examples of this type are \(^{96}\)Zr and \(^{116}\)Cd shown 
in Fig. \ref{fig:cumsumL-exc1}, panels (c) and (e). Characteristic feature of type 2
is that a large number of intermediate states give important contributions to the NMEs. 
In case of \(^{116}\)Cd, panel (e), around 50\% of the total NME comes from transitions through
the ground state of the intermediate nucleus. The other 50\% is distributed rather evenly
on the interval 0-20 MeV.

Using the multipole decompositions, we extracted the most important
multipole components contibuting to the light neutrino mediated 
\(0^+_\mathrm{gs} \rightarrow 0^+_{1} \) decay transitions. These most important components
were then again divided into contributions coming from different energy levels
of the \(0\nu\beta\beta\) intermediate nucleus. These contributions
are collected into Table \ref{tab:exc1-1} for \(A = 76-116 \) systems and into
Table \ref{tab:exc1-2} for \(A = 124-136 \) systems. Again we notice that often only
a few intermediate states give the largest contibution to the
dominant multipoles \(1^+\) and \(0^+\). Extreme case is the nucleus \(^{116}\)Cd
for which the dominant intermediate ground state gives 81\% of the
total \(1^+\) strength. Combining this with the fact that \(1^+\) is by far
the largest multipole component, we get a rather good approximation for the total 
NME by considering just a single virtual transition through the
\(1^+\) ground state of the intermediate nucleus \(^{116}\)In.
As for the ground-state-to-ground-state decays in some cases notable contributions are 
coming from high excitation energies, well above 10 MeV. 
There are high-energy contributions in case of $1^+$ multipole for all nuclei, and in the cases 
of $2^-$ and $2^+$ multipoles for \(^{130}\)Te and \(^{136}\)Xe.

\begin{table*}[]
\caption{Most important multipoles and intermediate states contributing
to the ground-state-to-excited-state \(0\nu\beta\beta\) decays mediated by
the light neutrino exchange. Columns \(E\) give the energies (in MeVs) and multipoles of the
intermediate states. Multipoles are organized from left to right in terms of their
importance, the most important being on the left. Columns labeled C give the corresponding NME contributions.
Last two numbers in each C column give the summed contribution and the percentual
part which the displayed states give to the total multipole strength.}
\label{tab:exc1-1}
\begin{ruledtabular}       
\begin{tabular}{lcccccccccc}
Nucleus & \(E(1^+)\) &  C  & \(E(0^+)\) &  C  & Nucleus &  \(E(1^+)\)  & C  
&  \(E(0^+)\)  &  C &    \\
\(^{76}\)Ge&	0.00	&	0.390		&	2.86	&	0.053	
& \(^{82}\)Se	&	0.19	&	0.149	&	5.45	&	0.471	\\
&	2.95	&	0.062		&	4.28	&	1.204	&	
&	5.32	&	0.179	&		&		\\
&	4.94	&	0.501		&		&		&	
&	7.01	&	0.145	&		&		\\
&	7.87	&	0.435		&		&		&	
&	14.53	&	0.274	&		&		\\
&	13.52	&	0.522		&		&		&	
&		&		&		&		\\
&		&	1.909		&		&	1.256	&	
&		&	0.746	&		&	0.471	\\
&		&	92\%		&		&	99\%	&	
&		&	90\%	&		&	96\%	\\
\hline
Nucleus & \(E(1^+)\) &  C  & \(E(2^-)\) &  C  & \(E(2^+)\) &  C  & \(E(0^+)\) &  C  
&  \(E(1^-)\) & C   \\  
\(^{96}\)Zr&	0.00	&	0.700	&	0.92	&	0.283	&	0.64	
&	0.122	&	5.37	&	0.033	&	2.28	&	0.032	\\
&	7.81	&	0.051	&	9.66	&	0.057	&	1.63	
&	0.180	&	6.64	&	0.025	&	2.52	&	0.080	\\
&	8.99	&	0.058	&		&		&	5.37	
&	0.035	&	7.32	&	0.037	&	4.46	&	0.026	\\
&	11.65	&	0.065	&		&		&	6.92	
&	0.027	&	8.02	&	0.036	&	5.27	&	0.068	\\
&	16.17	&	0.354	&		&		&		
&		&	10.12	&	0.396	&	11.33	&	0.026	\\
&		&		&		&		&		
&		&	12.00	&	0.013	&		&		\\
&		&	1.228	&		&	0.340	&		
&	0.364	&		&	0.540	&		&	0.232	\\
&		&	91\%	&		&	60\%	&		
&	65\%	&		&	98\%	&		&	59\%	\\
\hline
Nucleus & \(E(1^+)\) &  C  & \(E(0^+)\) &  C  & Nucleus &  \(E(1^+)\)  & C  
&  \(E(0^+)\)  &  C &    \\
\(^{100}\)Mo&	9.42	&	0.090	&	9.01	&	0.076	
& \(^{110}\)Pd	&	0.00	&	0.308	&	10.26	&	0.165	\\
&	13.62	&	0.079	&	9.55	&	0.150	&	
&	10.22	&	0.059	&		&		\\
&	14.31	&	0.072	&		&		&	
&	13.73	&	0.083	&		&		\\
&	15.12	&	0.063	&		&		&	
&	14.56	&	0.055	&		&		\\
&		&		&		&		&	
&	15.16	&	0.072	&		&		\\
&		&	0.303	&		&	0.226	&	
&		&	0.577	&		&	0.165	\\
&		&	65\%	&		&	77\%	&	
&		&	76\%	&		&	70\%	\\
\hline
Nucleus & \(E(1^+)\) &  C  &  &    & &    &   &    &   &    \\ 
\(^{116}\)Cd&	0.00	&	0.735	&	&	&	&	
&	&	&	&	\\
&		&	0.735	&	&	&	&	&	
&	&	&	\\
&		&	81\%	&	&	&	&	
&	&	&	&	\\
\end{tabular}
\end{ruledtabular}
\end{table*}
\begin{table*}[]
\caption{Most important multipoles and intermediate states contributing
to the ground-state-to-excited-state \(0\nu\beta\beta\) decays mediated by
the light neutrino exchange. Columns \(E\) give the energies (in MeVs) and multipoles of the
intermediate states. Multipoles are organized from left to right in terms of their
importance, the most important being on the left. Columns labeled C give the corresponding NME contributions.
Last two numbers in each C column give the summed contribution and the percentual
part which the displayed states give to the total multipole strength.}
\label{tab:exc1-2}
\begin{ruledtabular}         
\begin{tabular}{lcccccccccc}
Nucleus & \(E(1^+)\) &  C  & \(E(0^+)\) &  C  &  &    &   &    &   &    \\
\(^{124}\)Sn&	0.00	&	0.101	&	2.70	&	0.667	&	
&	&	&	&	&		\\
&	0.66	&	0.433	&	4.60	&	0.088	&	&	
&	&	&	&		\\
&	1.00	&	0.146	&	7.35	&	0.567	&	&	
&	&	&	&		\\
&	2.25	&	0.051	&		&		&	&	
&	&	&	&		\\
&	2.56	&	0.227	&		&		&	&	
&	&	&	&		\\
&	3.31	&	0.664	&		&		&	&	
&	&	&	&		\\
&	6.72	&	0.289	&		&		&	&	
&	&	&	&		\\
&	7.91	&	0.424	&		&		&	&	
&	&	&	&		\\
&	13.09	&	0.473	&		&		&	&	
&	&	&	&		\\
&	13.86	&	0.161	&		&		&	&	
&	&	&	&		\\
&		&	2.968	&		&	1.322	&	&	
&	&	&	&		\\
&		&	94\%	&		&	94\%	&	&	
&	&	&	&		\\
\hline
Nucleus & \(E(1^+)\) &  C  & \(E(0^+)\) &  C  & \(E(2^-)\) &  C  & \(E(2^+)\) 
&  C  &   &    \\ 
\(^{130}\)Te&	0.25	&	0.390	&	7.40	&	0.719	
&	0.97	&	0.340	&	0.41	&	0.019	&	&	\\
&	1.50	&	0.082	&	8.35	&	0.631	&	5.07	
&	0.073	&	0.59	&	0.047	&	&	\\
&	2.32	&	0.097	&		&		&	17.26	
&	0.106	&	1.83	&	0.017	&	&	\\
&	2.76	&	0.134	&		&		&	18.95	
&	0.083	&	2.80	&	0.043	&	&	\\
&	4.53	&	0.064	&		&		&		
&		&	4.70	&	0.024	&	&	\\
&	5.59	&	0.338	&		&		&		
&		&	4.86	&	0.027	&	&	\\
&	7.57	&	0.592	&		&		&		
&		&	5.18	&	0.033	&	&	\\
&	14.66	&	0.065	&		&		&		
&		&	6.27	&	0.016	&	&	\\
&	15.01	&	0.259	&		&		&		
&		&	6.75	&	0.051	&	&	\\
&	15.07	&	0.773	&		&		&		
&		&	9.10	&	0.032	&	&	\\
&	15.39	&	0.089	&		&		&		
&		&	10.57	&	0.045	&	&	\\
&		&		&		&		&		
&		&	17.87	&	0.026	&	&	\\
&		&		&		&		&		
&		&	23.14	&	0.024	&	&	\\
&		&	2.882	&		&	1.351	&		
&	0.602	&		&	0.390	&	&	\\
&		&	95\%	&		&	99.6\%	&		
&	79\%	&		&	65\%	&	&	\\
\hline
Nucleus & \(E(1^+)\) &  C  & \(E(0^+)\) &  C  & \(E(2^-)\) &  C  & \(E(2^+)\) &  C  &   &    \\ 
\(^{136}\)Xe&	2.30	&	0.366	&	12.24	&	0.317	
&	2.50	&	0.060	&	4.86	&	0.204	&	&	\\
&	3.06	&	0.050	&	12.54	&	0.712	&	4.86	
&	0.056	&	18.58	&	0.083	&	&	\\
&	3.75	&	0.066	&		&		&	5.29	
&	0.113	&	20.34	&	0.059	&	&	\\
&	4.46	&	0.226	&		&		&	8.34	
&	0.053	&		&		&	&	\\
&	10.12	&	0.232	&		&		&	14.11	
&	0.054	&		&		&	&	\\
&	11.52	&	0.565	&		&		&		
&		&		&		&	&	\\
&	17.12	&	0.137	&		&		&		
&		&		&		&	&	\\
&	17.78	&	1.031	&		&		&		
&		&		&		&	&	\\
&		&	2.675	&		&	1.029	&		
&	0.335	&		&	0.346	&	&	\\
&		&	93\%	&		&	101\%	&		
&	53\%	&		&	67\%	&	&	\\
\end{tabular}
\end{ruledtabular}
\end{table*}

\subsection{conclusions}

In this article we have extended our previous works \cite{Hyvarinen2015,Hyvarinen2016} 
on the ground-state-to-ground-state and ground-state-to-excited-state
\(0\nu\beta\beta\) decay transitions.
In the present work we have concentrated our studies on the intermediate contributions
to the NMEs involved in the light-neutrino mediated \(0\nu\beta\beta\) decay. 
We have calculated the intermediate state multipole 
decompositions of the NMEs and extracted the most important multipole components.
Cumulative sums of the NMEs were calculated to investigate the important energy
regions contributing to the \(0\nu\beta\beta\) transitions.
Finally, the most important multipole components were divided into contributions
coming from the virtual transitions through the individual states of the 
\(0\nu\beta\beta\) intermediate nuclei. An extensive tabulation of these
important intermediate states were given for all the nuclei considered in this paper. 

We have done these computatations
by using realistic two-body interactions and single-particle bases. All the appropriate 
short-range correlations, nucleon form factors and higher-order nucleonic weak currents
are included in our present results. 

We found in the calculations that often there exists only a few relevant
intermediate states which collect most of the strength corresponding
to a given multipole. We also found that there exists a single-state
dominance in the important \(2^-\) components related to the ground-state decays
of nuclei \(^{76}\)Ge, \(^{82}\)Se and perhaps also for \(^{96}\)Zr.

\section*{Acknowledgments}\label{thanks}

This work has been partially supported by the Academy of Finland under 
the Finnish Centre of Excellence Programme 2012-2017 (Nuclear and 
Accelerator Based Programme at JYFL).

\end{document}